# OPERATION OF THE H− LINAC AT FNAL[*]

K. Seiya [†], T. Butler, D. Jones, V. Kapin, K. Hartman, S. Moua, J. -F. Ostiguy, R. Ridgway,
R. Sharankova, B. Stanzil, C. Y. Tan, J. Walters, M. Wesley
Fermi National Accelerator Laboratory, Batavia 60510 IL, USA
M. Mwaniki
Illinois Institute of Technology, Chicago, IL 60616, USA

## Abstract

The Fermi National Accelerator Laboratory (FNAL) Linac has been in operation for 52 years. In approximately four years, it will be replaced by a new 800 MeV superconducting machine, the PIP-II SRF Linac. In the current configuration, the Linac delivers H− ions at 400 MeV and injects protons by charge exchange into the Booster synchrotron. Despite its age, the Linac is the most stable accelerator in the FNAL complex, reliably sending 22 mA in daily operations. We will discuss the status of the operation, beam studies, and plans.

## INTRODUCTION

FNAL is leading the intensity frontier by providing high intensity proton beams to high energy experiments. The Linac delivers H− ions at 400 MeV to inject protons by charge exchange into the Booster, a 15 Hz rapid cycling booster synchrotron.

H− beam is supplied to the Linac by an RFQ injection line (RIL) which consists of a magnetron ion source, a low energy beam transport (LEBT), a radio frequency quadrupole (RFQ) with acceleration from 35 to 750 keV and a medium energy beam transport (MEBT) with a buncher cavity [1]. The Linac is divided into two sections: (1) a 201.25 MHz Drift Tube Linac (DTL) where the H− beam is accelerated to 116 MeV [2] and (2) a 805 MHz Side Coupled Linac (SCL) which further accelerates the beam to 400 MeV [3].

The Linac has been in operation for 52 years and reliably sends 22 mA in daily operations. Our present goals are:
- Minimizing machine downtime and providing stable beam to users.
- Increasing the output current to more than 30 mA.
- Using Machine Learning (ML) to optimize RF parameters and automate machine tuning.

In this paper, we present the history and status of the beam operation and machine studies for the last 5 years.

## STATUS OF OPERATION

The Proton Improvement Plan (PIP) aimed to run 4.3E12 protons per pulse in the Booster at 15 Hz and successfully accomplished its goal in 2017 [4]. As a part of PIP, modulators for the DTL RF system were upgraded for more reliable operations [5]. With 24/7 user support as our first priority, the Linac has been providing 22 mA consistently with 96 % machine up-time for the last 5 years.

### Beam current and efficiency

The beam current in the LEBT, upstream of the Linac, and end of the Linac at 400 MeV are shown in Fig. 1. The transmission efficiency in RIL has been 40 % and Linac 92 %. In 2017, a collimator with an aperture size of 9.9 mm × 14.5 mm was installed in front of the Linac in order to minimize beam loss in the Booster. The average RMS beam sizes are ±2.7 mm horizontal and ±3.5 mm vertical.

Two major machine failures occurred in the last 5 years. A water leak damaged a quadrupole magnet in a drift tube (DT) in Tank 5 and caused an intermittent short on the wire around the quadrupole magnet. Replacing the exposed wires on the tank and isolating the quadrupole magnet from the ground took 72 hrs. An end plate on Tank 4 was missing one push screw, which was supposed to push the plate to the flange for a good RF contact, thus creating a gap which caused a spark in the cavity. A comparison between the three gradient detector signals at low, mid, and high energies in the tank indicated that the spark occurred at lower energy. Finding the source and completing the repair took 112 hrs.

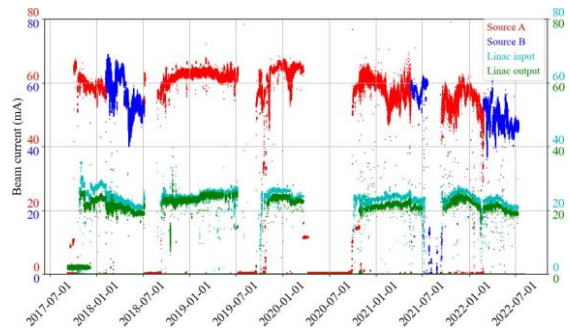

Figure 1: Linac beam current over last 5 years.

### Drift tube replacement and alignment

**Rebuilding DT**  The DTL has 207 DTs in Tank 1 - Tank 5. Each DT was built with oxygen free high conductivity copper and contains a quadrupole magnet that isolates it from a vacuum. The length varies from 48 mm to 409 mm throughout the Linac, and the bore size varies in different tanks. Twelve DT failures have been recorded since 1986 and four DTs were replaced in the last 4 years in Tank 2 – 5 due to sparking, vacuum leak, or either a short or open magnet circuit. Some spare DT bodies without end caps were built before 1970 and one of them was used for the DT in Tank 3 by adjusting the length with extension rings and end caps which were welded with an electron beam. Since there were no suitable spares for the drift tubes for Tank 2 and Tank 4, two long and short drift tubes which had been in a showcase at the Linac gallery were used by cutting off their end caps. The drift tube for Tank 5 was built from scratch by brazing a stem to a cylinder. All drift tubes were built within 0.1 mm accuracy in their length.

[*] Work supported by Fermilab Research Alliance, LLC under Contract No. DE-AC02-07CH11359 with the United States Department of Energy
[†] kiyomi@fnal.gov

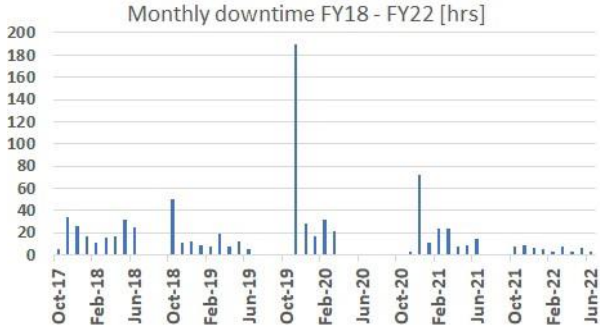

Figure 2: Monthly downtime over last 5 years.

Quadrupole magnets are water cooled indirectly by contact with the drift tube body. The gap between the drift tube body and magnet was minimized to less than 0.2 mm with 0.1 mm on average for heat transfer. The magnet temperature was measured at four different cooling conditions where the LCW temperature was 35 °C (with flow rates of 1.4 and 0.2 gpm), 18.3 °C, and air cooling. The magnet temperature reached equilibrium at 30 °C above the LCW temperature for the three cases with water cooling. The heat flow rate $Q$ is given by Eq. 1 where $k$ is the thermal conductivity of the air, $L$ and $r$ are the magnet's length and radius, $\Delta T$ is the temperature difference, and $\Delta r$ is the air gap length between the quadrupole magnet and the DT body.

$$Q = 2\pi k L (\Delta T)/\ln((r + \Delta r)/r) = 287 \text{ W} \qquad (1)$$

The heat flow rate is nearly equal to the input power of 258 W calculated from the current on a pulse power supply which sends a 250 A, 272 Hz sinusoidal current at 15 Hz.

All DT quadrupoles were tested with 250 A at 15 Hz for 2 months before installation.

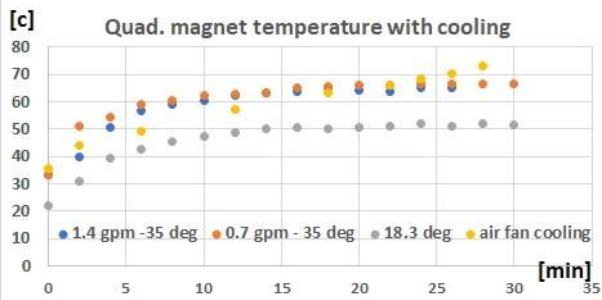

Figure 3: Quadrupole temperature with cooling.

**DT alignment** Four drift tubes were replaced during the summer shutdowns in 2018 – 2021. A survey was conducted before and after the replacement by placing an alignment fixture with a ball-mounted hollow reflector for a laser tracker upstream and downstream of each DT. The accuracy of the position was measured to within 0.13 mm. The drift tubes were aligned within 0.25 mm accuracy on the best fit line from upstream to downstream of the DTs. A survey was conducted for all DTs in Tank 2 – Tank 5 over the last four years and is shown in Fig. 4. The figure also shows the positions of the tank centers estimated from control points outside of each tank. The vertical position of the drift tubes has a maximum offset of 3.5 mm. There is a 2 mm jump at the beginning of the SCL.

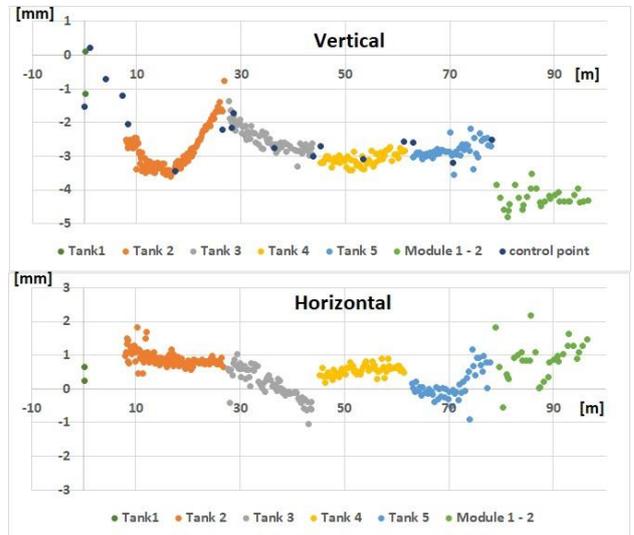

Figure 4: Survey for DTs in Tank 2 - 5 and Module 0 - 2.

## BEAM LOSS IN THE RFQ INJECTOR LINE

The transmission efficiency has to be improved in the RIL to achieve 30 mA output from the linac. All elements except the LEBT trim dipoles in the RIL were realigned to the original reference line (the Tank 1-Brass Line Frame) in 2019, and realigned to the line that followed the control points in Tank 1 (Fig. 5) by tilting it by 0.25 mrad vertically in 2020. However there were no significant improvement in the transmission efficiency with this work.

An additional diagnostic which included a toroid, BPM and halo monitor was temporary installed between the MEBT and RFQ for two weeks of beam studies in the 2021 summer shutdown and found that 60 % of the beam was lost upstream of the MEBT. Measurements in a test stand indicated that the emittances are too large compared to the RFQ acceptance and significant loss occurred in the LEBT [6].

## IMPLEMENTING MACHINE LEARNING INTO OPERATION

The compact nature of the Linac drift-tube structure does not provide sufficient space to accommodate extensive beam diagnostics. As a consequence, machine tuning has to rely on combining a limited set of diagnostics with simulations and is unlikely to be optimal. External factors such as ambient temperature and humidity variations are known to affect resonant frequencies. In addition, for a variety of complex causes, the energy and phase space distribution of particles emerging from the ion source are subject to fluctuations. Recent developments suggest that global and dynamic techniques based on machine learning should make it possible to overcome these obstacles.

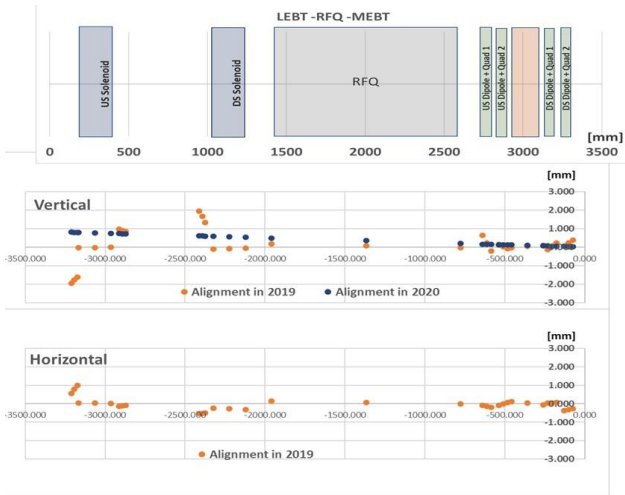

Figure 5: RIL alignment in 2019 and 2020.

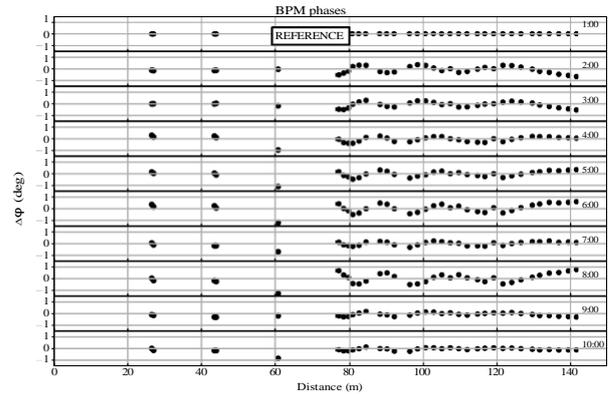

Figure 6: Drift of longitudinal positions in the Linac.

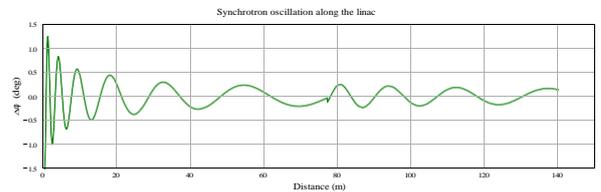

Figure 7: Simulated synchrotron oscillations in the Linac.

*Daily Linac tuning*

Operators typically tune the Linac twice a day by changing three RF phases to minimize total beam loss and maximize the output current. Fig. 6 shows the drift of the longitudinal position measured on BPM phase signal for 10 hrs without any parameter changes. The phase drifts are known to depend on the Linac gallery temperature or extracted beam energy from the ion source, etc., however the details are not well understood.

Assuming a 2 degree phase offset from the synchronous particle at the beginning of the Tank 1, phase oscillation of the synchrotron motion was calculated in Fig. 7 with a simple acceleration model (a single kick for each gap) using the design structure of the $2\pi$ mode DTL and $\pi/2$ mode SCL. There are not enough diagnostics to measure the $\sim 10$ synchrotron oscillations along the Linac.

*RF parameter optimization with machine learning*

We are exploring ML applications for RF parameter optimization [7]. Our approach has been to train a Deep Neural Network (DNN) to identify underlying correlations between observed diagnostic data and Linac RF parameters. The plan is to use the network predictions as part of a control scheme that adjusts phase set points to restore the Linac back to a desired state (as defined by diagnostic readings). The choice of DNN among all available ML algorithms was driven by the fact that in principle a network with enough nodes can approximate almost any analytical relation, without the need to resort to exact information about underlying correlations between observables. As a first attempt, we trained a network to predict changes in RF phase set points for 3 cavities (RFQ, MEBT Buncher and Tank 5) given BPM data. Fig. 8 shows network predictions compared to true settings. Initial results were very promising; performance on par with human operator tuning had been demonstrated in specific instances.

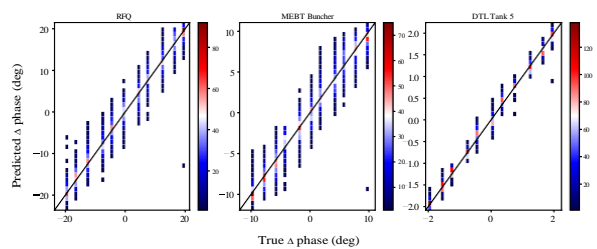

Figure 8: DNN predicted RF phase change vs. truth.

## SUMMARY

The Linac has been delivering 22 mA H$^-$ beam to the Booster with 96 % uptime for the past 5 years. Two major machine failures have occurred which took more than 50 hrs to fix. Four DTs were replaced in the low energy tanks. To improve beam transmission, alignment, beam studies, beam simulation, and instrumentation upgrade have been conducted in the RIL and Linac. RF optimization using ML has been studied and appears promising. All efforts will be continued to achieve the Linac performance goals.

# REFERENCES

[1] C. Y. Tan et al., "The 750 keV RFQ Injector Upgrade", in *FNAL Beams-Doc-3646-V16*, 2013.

[2] G. W. Wheeler, "The Brookhaven 200-MeV Proton Linear Accelerator", in *Particle Accelerators*, 1990, Vol. 9, pp 1–156.

[3] R. Noble, "The Fermilab Linac Upgrade", in *Proc. LINAC'90*, Albuquerque, NM, USA, 1990, pp. 26–30.

[4] W. Pellico et al., "FNAL - The Proton Improvement Plan (PIP)", in *Proc. IPAC14'19*, Dresden, Germany, June 2014.

[5] T. A. Butler et al., "Development of a Marx Modulator for FNAL Linac", in *Proc. NAPAC'19*, East Lansing, MI, USA, September 2019.

[6] D. C. Jones et al., "Investigation of the Beam Propagation through the FNAL LEBT", in *Proc. LINAC2022*, Liverpool, UK, September 2022.

[7] R. Sharankova et al., "Diagnostics for Linac Optimization with Machine Learning", in *Proc. NAPAC'22*, Albuquerque, NM, USA, August 2022.